# Dalle opere orfane, un nuovo ruolo delle biblioteche per il pubblico dominio e l'utilità sociale

Federico Leva[1]

## Sommario

Sintetizzando le nuove norme sulle opere orfane, mostriamo come esse affermino la prevalenza dell'interesse pubblico e consentano a biblioteche e altri enti beneficiari di migliorare i propri servizi. Sosteniamo quindi che questi si inquadrano nella loro missione nei confronti del pubblico dominio e sono un primo passo per la sua completa realizzazione, mediante il lavoro di ciascuno e la riforma del diritto d'autore europeo. In caso contrario, la cultura europea sparirà.

## Abstract

Summarising the new orphan works law of Italy, we show how it makes the public interest prevail and allows libraries and other beneficiaries to improve their services. We then argue that such services are part of their mission towards the public domain and are a first step for its complete accomplishment, by the work of each and a reform of european copyright. Failing that, European culture will disappear.

## Parole chiave

copyright; public domain; digital libraries; european union; directive 2012/28/eu; orphan works; hostage works; mass-digitization; Google Books; Internet Archive

## Introduzione

Nel contesto della modernizzazione digitale della società europea e in specie dei suoi servizi culturali, la digitalizzazione di massa è spesso assurta a simbolo del progresso e delle sue difficoltà. Fra i molti ostacoli e storture, è stato esemplare il problema delle opere orfane (opere i cui titolari sono "spariti"), che infatti è stato oggetto di una direttiva finalmente recepita in Italia nel novembre 2014.[2] Per maggiore contesto, si rimanda a *Digitalia* 2013/2[3] e altri;[4] [5] qui intendiamo concentrarci sulle prospettive poste dai recenti

---

1. Associazione Wikimedia Italia (socio). Testo pubblicato con licenza Creative Commons Attribuzione 4.0 Internazionale, <https://creativecommons.org/licenses/by/4.0/>. L'autore ringrazia Rosa Maiello (Università degli Studi di Napoli "Parthenope") per i preziosi commenti utili a dissipare imprecisioni e fraintesi.
2. D.Lgs. 2014-11-10, n. 163. *Attuazione della direttiva europea 2012/28/UE su taluni utilizzi consentiti di opere orfane*. <http://www.normattiva.it/uri-res/N2Ls?urn:nir:stato:decreto.legislativo:2014-11-10;163!vig=2014-11-25>. Integrazione della legge 1941-04-22, n. 633 (articoli da 69-bis a 69-septies). <http://www.normattiva.it/uri-res/N2Ls?urn:nir:stato:legge:1941-04-22;633!vig=2014-11-25>.
3. Rosa Maiello. *Politiche e legislazione dell'Unione Europea per la digitalizzazione del patrimonio culturale*. «Digitalia», 2013/2. <http://digitalia.sbn.it/article/view/822>.
4. Maria Cassella. *Rights management in digitization projects: public domain and orphan works*. «JLIS.it», [S.l.], v. 4, n. 2, p. 223-254, jul. 2013. ISSN 2038-1026. <http://leo.cineca.it/index.php/jlis/article/view/8797>. doi:10.4403/jlis.it-8797.
5. Antonella De Robbio. *La Gestione dei Diritti nelle digitalizzazioni di massa: un'analisi alla luce del caso Google Book Search*. «Bibliotime» 12.2 (2009). <http://www.aib.it/aib/sezioni/emr/bibtime/num-xii-2/derobbio.htm>.

aggiornamenti.

Il fatto che ci si ponga il problema delle opere orfane è indice di un successo moderno, cioè la possibilità emergente di mettere vaste masse di opere, prima impossibili a distribuirsi, a disposizione del pubblico e dei creativi contemporanei (due categorie peraltro sempre piú sovrapposte[6]). D'altro canto, è indice del palese fallimento della normativa sul diritto d'autore: i diritti della collettività sono ristretti in favore di monopolî privati sulla distribuzione delle opere culturali;[7] nel momento in cui la stragrande maggioranza dei titolari non è in grado di utilizzare tali diritti, è evidente che ne resta solo il danno all'utilità sociale e alla cultura collettiva, ciò che contraddice le premesse del diritto d'autore stesso nonché dell'art. 41 della Costituzione.[8] Normando l'uso delle opere orfane, si è riconosciuta questa contraddizione e si è affermata la prevalenza del bene comune, a nostro avviso sintetizzata dal principio che «il pubblico dominio è la regola, il copyright è l'eccezione» (*Manifesto del pubblico dominio*; sull'argomento, si veda oltre).

*Illustrazione 1: Europeana, testi in pubblico dominio: pochi paesi svettano, Italia al 16º posto*

---

6 È facile fare ancora una volta l'esempio di Wikipedia e dei progetti Wikimedia: opere collettive che, nelle loro oltre 800 edizioni linguistiche, hanno trasformato oltre 15 milioni di lettori in autori, di cui 5 fra i soli utenti registrati di Wikipedia in inglese. https://stats.wikimedia.org/IT/TablesWikipediaEN.htm#editdistribution
7 Christopher M. Toula – Gregory C. Lisby. *Towards an affirmative public domain*. «Cultural Studies», Volume 28, Issue 5-6, 2014. DOI: 10.1080/09502386.2014.886490
8 https://it.wikisource.org/wiki/Italia,_Repubblica_-_Costituzione#Art._41

# Il contenuto della norma

Veniamo dunque al contenuto della norma.

- Essa afferma, al primo articolo, che i beneficiari[9] possono, concretamente: riprodurre le opere orfane in qualunque modo; metterle a disposizione degli utenti per un accesso illimitato anche nel world wide web. Eventuali ricavi devono andare a sostenere tali costi.

- Procede poi a specificare a quali opere a stampa o fonogrammi, di origine europea, si applichi la norma; come attuare la ricerca diligente dopo cui un'opera può essere dichiarata orfana; e come i titolari dei diritti possano, "ricomparendo", porre fine allo stato di opera orfana ed eventualmente ottenere un compenso.

- Infine, un lungo articolo di due pagine fornisce un elenco *non esaustivo* di ben 24 fonti consultabili per assolvere alla ricerca diligente (in Italia), indicando però esplicitamente VIAF e ARROW come modalità di consultazione delle varie fonti per i libri pubblicati.

È importante sottolineare che i titolari ricomparsi avranno diritto a un "equo compenso" e per ottenerne uno maggiore saranno costretti a un tentativo di conciliazione, ma questo non rende i beneficiari immuni da ogni eventuale causa giudiziaria.

Quanto ai dettagli della ricerca diligente, in sede di parere consultivo[10] una commissione parlamentare ha obiettato che non è sufficientemente specificata: alquanto assurdo, quando si sono introdotti in legge diversi riferimenti a strumenti tecnologici specifici (spesso privati) la cui esistenza di lungo periodo non è garantita dalla legge stessa.[11] Il problema è semmai che l'elenco, dovendo essere specifico per risolvere il problema europeo pluridecennale di che cosa costituisca "ricerca diligente", non ha potuto essere dichiarato esaustivo. Quindi, un ente potrebbe essere accusato di insufficiente diligenza pur avendo consultato tutte le fonti indicate dalla legge.

Questi due aspetti, però, discendono dalla discussione in sede europea e quindi sono già stati trattati in passato: da verificare è piuttosto l'uso reale che se ne farà; si veda "Attuazione" più avanti.

# La preminenza dell'interesse pubblico...

Particolarmente interessante è, a nostro avviso, riflettere sul primo articolo (ora articolo 69-bis della legge 633/1941).

Innanzi tutto, l'eccezione si applica esclusivamente a «biblioteche, gli istituti di istruzione e i musei, accessibili al pubblico, nonché gli archivi, gli istituti per il patrimonio cinematografico o sonoro e le emittenti di servizio pubblico». È quindi di natura chiaramente diversa da altre eccezioni quali il diritto di corta citazione, che è ammesso in quanto ha fini illustrativi e non compete con un uso commerciale dell'opera. Invece, l'uso delle opere orfane è consentito in funzione delle specifiche caratteristiche di questi enti, che nonostante l'elencazione risultano alquanto flessibili: per esempio, le biblioteche sono incluse qualora "accessibili al pubblico" a prescindere che siano enti di diritto pubblico.

---

9   Più avanti chiamati semplicemente "beneficiari" o anche solo "biblioteche" per chiarezza espositiva, senza alcuna volontà di esclusione degli altri importanti casi.
10  La legge sul diritto d'autore è infatti stata emendata con un decreto legislativo del governo, sulla base della vastissima legge di delegazione europea 2013. http://www.normattiva.it/uri-res/N2Ls?urn:nir:stato:legge:2013-08-06;96 Il parlamento ha quindi ancora una volta mancato ai propri doveri di riforma del diritto d'autore ed è stato solo consultato in merito allo schema di decreto predisposto dal governo.
11  Riferimenti a fonti e strumenti specifici non sono invece presenti nella direttiva, ma solo in un allegato.

Similmente, i commi 2 e 3 stabiliscono che «Le opere orfane possono essere utilizzate dalle organizzazioni di cui al comma 1 unicamente per scopi connessi alla loro missione di interesse pubblico [...]. I ricavi eventualmente generati [...] sono impiegati per coprire i costi per la digitalizzazione delle opere orfane e per la messa a disposizione del pubblico delle stesse». Opportunamente, non vengono stabiliti in dettaglio a priori gli utilizzi possibili e, in particolare, non si tratta esplicitamente l'uso commerciale (che a nostro avviso è sempre un falso problema[12]): un uso è ammesso se persegue la missione dell'ente e tutti i ricavi sono impiegati per le opere orfane. Per qualsiasi ente pubblico o privato non a scopo di lucro che abbia fra i propri scopi il mettere opere culturali a disposizione del pubblico, non sembrerebbe difficile rispettare questi requisiti: "basta" fare il proprio mestiere e spendere oculatamente.

Mentre i beneficiari diretti sono pochissimi, almeno le eccezioni loro concesse sono flessibili e sensate. Siamo molto lontani dal *fair use* statunitense, che si applica a chiunque ed è talmente vasto da aver consentito, dopo lungo travaglio, una prima conferma da parte del giudice Chin della legalità della digitalizzazione e uso da parte di Google di decine di milioni di opere.[13] Ma è comunque un sollievo, dopo le formule a cui il legislatore italiano ci ha abituato, quali il famigerato «a bassa risoluzione o degradate, per uso didattico o scientifico e solo nel caso in cui tale utilizzo non sia a scopo di lucro» dell'articolo 70.1-bis introdotto nel 2008 e per giunta mai attuato.[14]

## ...e solo pubblico

Ancora piú importante, però, è a nostro avviso il comma 5, che vale la pena di citare integralmente:

> Le organizzazioni di cui al comma 1, nell'adempimento della propria missione di interesse pubblico, hanno la facoltà di concludere accordi volti alla valorizzazione e fruizione delle opere orfane attraverso gli utilizzi di cui al comma 1. Tali accordi non possono imporre ai beneficiari dell'eccezione di cui al presente articolo alcuna restrizione sull'utilizzo di opere orfane e non possono conferire alla controparte contrattuale alcun diritto di utilizzazione delle opere orfane o di controllo dell'utilizzo da parte dei beneficiari. Gli accordi non devono essere in contrasto con lo sfruttamento normale delle opere, né arrecare un ingiustificato pregiudizio agli interessi dei titolari dei diritti.[15]

Il comma 5 afferma ulteriormente che ai beneficiari è concesso di utilizzare le opere orfane, ma non per sé stessi, bensí solo per la collettività. I diritti di utilizzo possono anche essere sfruttati in collaborazione con altri enti non beneficiari, ma devono restare integralmente nel perimetro dell'interesse pubblico dei beneficiari; non possono essere alienati o ridotti, quindi la controparte contrattuale sarà – per cosí dire – sempre subordinata e dipendente.

---

12  Paul Klimpel. *La conoscenza libera basata sulle licenze Creative Commons: conseguenze, rischi ed effetti collaterali della clausola "Non Commerciale – NC"*. 2012. https://meta.wikimedia.org/wiki/Free_knowledge_based_on_Creative_Commons_licenses/it
13  Corynne McSherry. *Court Upholds Legality of Google Books: Tremendous Victory for Fair Use and the Public Interest*. EFF, 2013-11-14. https://www.eff.org/deeplinks/2013/11/court-upholds-legality-google-books-tremendous-victory-fair-use-and-public
14  http://www.interlex.it/testi/l41_633.htm#70
15  L'ultima frase è ridondante, in quanto riproduce le previsioni dell'articolo 71-nonies.

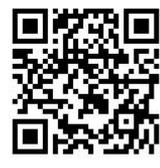

*Illustrazione 2: Pagina, collegata dall'OPAC SBN, di un'opera BNCR in Google*

È piú facile comprendere l'importanza di questo punto guardando a un noto esempio, quello della collaborazione fra Google e il MIBACT (le due biblioteche nazionali centrali),[16] di cui è stato resto noto il testo contrattuale.[17] Il progetto in questione riguarda esclusivamente opere in pubblico dominio e quindi non è perfettamente rappresentativo delle problematiche relative alle opere orfane, in particolare non comporta cessione di diritti di utilizzazione che Google già non avesse; però ci sembra utile a illustrare il contesto.

L'accordo in questione non ha effetti su terze parti né cede diritti d'autore (art. 4.1, 4.2), di cui nemmeno il Ministero può disporre; non comporta nemmeno un'esclusiva per i due contraenti (art. 2.7). Tuttavia, è piuttosto ambiguo nel parlare di una "piena titolarità delle Copie Digitali di Google"; ribadisce i diritti che Google (come tutti) ha sul pubblico dominio (art. 4.3), ma definisce in modo alquanto ristretto i diritti dei terzi. «[Google] provvederà a fornire un servizio gratuito agli Utenti Finali per l'accesso alla visualizzazione del testo integrale dell'opera riprodotta in tale Copia Digitale»; «Google non è tenuta a rendere disponibili una o tutte le Copie Digitali di Google mediante i propri Servizi». Fin qui, nulla di particolarmente sorprendente.

Il testo però prosegue: «il Ministero e/o la Biblioteca non potranno pubblicare né utilizzare in altro modo le Copie Digitali della Biblioteca al di fuori di quanto espressamente consentito dal presente Accordo» (art. 4.6); e si capisce finalmente lo scopo: «La Biblioteca avrà la facoltà di utilizzare le Copie Digitali [...] non potranno richiedere o ricevere il pagamento o altra remunerazione [...] dovrà provvedere [...] all'attivazione di misure tecnologiche [...] per limitare l'accesso automatizzato [...] non permettano, a terzi di (a) trasferire [...] (b) ridistribuire porzioni delle Copie Digitali della Biblioteca, o c) effettuare il trasferimento elettronico automatizzato e sistematico [...] garantire che porzioni sostanziali delle Copie Digitali della Biblioteca <u>non vengano trasferite [...] od in altro modo resi liberamente accessibili al pubblico</u>. [...] Tutte le restrizioni ed i requisiti [...] verranno meno allo scadere del quindicesimo anno» (art. 4.7; enfasi nostra).

---

16 MIBACT. *Progetto "Google Books"*. <http://www.librari.beniculturali.it/opencms/opencms/it/progetti/pagina_0002.html>.
17 *Accordo fra Google Books e il Ministero per i Beni e le Attività Culturali per la digitalizzazione delle opere delle biblioteche italiane*. 2010. <https://archive.org/details/AccordoMiBACGoogle>.

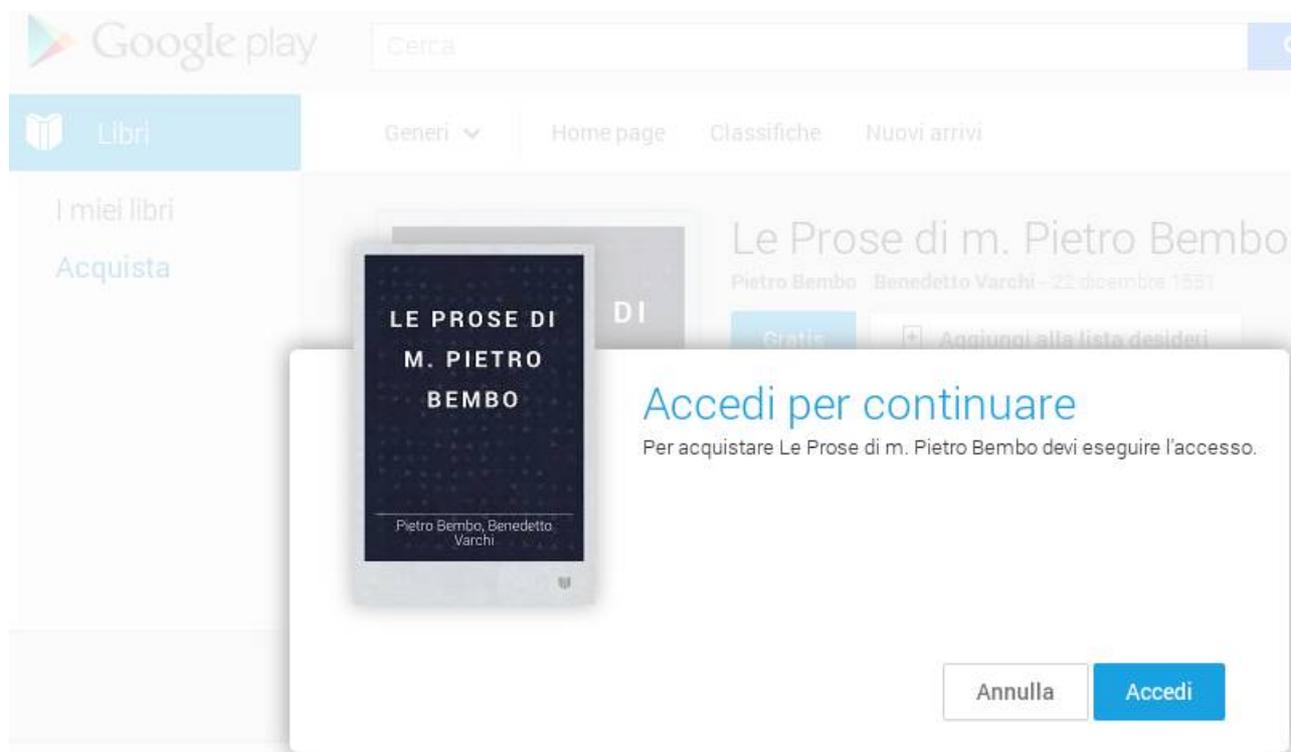

*Illustrazione 3: Google richiede di accedere per scaricare l'opera (in PDF)*

È evidente lo scopo: fino a qualche anno fa, per scaricare il PDF di un libro di Google (Ill. 2) era "sufficiente" compilare un Captcha e non essere lettori troppo voraci (pena il blocco perpetuo);[18] ora è invece necessario registrarsi e accedere a "Google Play", piattaforma ("market") che offre o vende una pletora di servizi (Ill. 3). Le clausole di cui sopra mirano a garantire che gli utenti di opere della Biblioteca siano costretti a diventare clienti registrati di Google, e soprattutto che non esista un servizio concorrente. La Biblioteca potrebbe offrire essa stessa tale servizio, quindi in teoria non sussiste un monopolio di Google; non risulta però che abbia intenzione di farlo, ed è molto difficile che sia in grado: costruire una piattaforma che fornisca le garanzie richieste da Google è molto costoso, inoltre gli utenti ne avrebbero un'utilità marginale (per quanto comunque considerevole), perché la Biblioteca sarebbe costretta a erogare il servizio in modalità omogenee a quelle dei siti Google.

L'esclusiva (parziale) è dunque l'*obiettivo*. Il *metodo* legale seguito per ottenerla nel caso delle opere di pubblico dominio, però, non è disponibile per le opere orfane, su cui Google non potrebbe acquisire pieni diritti di utilizzazione. Vediamo allora quali sarebbero le possibilità di manovra.

Le pretese di Google sono in parte fondate sulla presunta distinzione fra "opera" e sua "copia digitale", ove la seconda acquisisce restrizioni simili al copyright ma fuori dal copyright,[19] e non possono comunque restringere i diritti preesistenti di terzi. Tuttavia, paradossalmente, quelle opere potevano essere oggetto di qualsiasi contratto privato proprio perché sono nel pubblico dominio; non cosí le opere orfane, il cui utilizzo è regolato dalla legge e a nostro avviso non può essere limitato da alcun accordo, visto il sopra citato comma 5. Se anche si ammettesse che le copie digitali possano essere soggette a diritti distinti da quelli delle opere medesime, si tratterebbe quanto meno di opere derivate e quindi esse stesse soggette (almeno) ai medesimi diritti d'autore. È dunque escluso ogni «controllo dell'utilizzo da parte dei beneficiari», che possono rendere le opere orfane liberamente accessibili al pubblico nei propri siti come meglio credono, e nemmeno volendo possono rinunciare a questo diritto. Inoltre, «non possono conferire alla controparte contrattuale alcun diritto di utilizzazione delle opere orfane»: come si è visto, un uso limitato potrebbe rientrare fra i diritti di Google a prescindere, almeno nella legislazione statunitense che prevede il *fair use;* ma non tutti i diritti dei beneficiari dell'eccezione sarebbero disponibili. Per giunta, alla luce del comma 3, ogni ricavo dovrebbe essere reinvestito per l'uso delle opere orfane: ma come quantificare i ricavi generati da un'opera inserita nel *mare magnum* dei servizi Google, che hanno le caratteristiche di cui sopra?

---

18  Brewster Kahle. *Aaron Swartz memorial at the Internet Archive*. 2013-01-24.
    https://archive.org/details/AaronSwartzMemorialAtTheInternetArchive?start=4680
19  Severine Dusollier. *Scoping Study on Copyright and Related Rights and the Public Domain*. World Intellectual Property
    Organisation Publication, March 2011. <http://ssrn.com/abstract=2135208>.

Insomma, a noi sembra che la legge imponga una forte presa di responsabilità da parte dei beneficiari, che non potranno delegare ad altri lo svolgimento dei propri compiti di interesse pubblico. Collaborazioni sono possibili, ma dovranno essere fondate su basi molto diverse che in passato. In particolare, un'eventuale riedizione dell'accordo con Google per le opere orfane, ammesso che sia legalmente possibile, potrebbe avvenire solo a condizione che ampie porzioni dei privilegi prima concessi a Google siano eliminate e che siano rivoluzionate le modalità di accesso alle opere nei siti Google per risolvere la questione dei ricavi ottenuti dalle stesse. L'operazione appare molto difficile. Considerando poi le difficoltà e i tempi del progetto già in corso,[20] per le opere orfane è probabilmente meglio seguire binari paralleli, in contemporanea.

# Attuazione

La norma è molto dettagliata e immediatamente funzionante, ma la legge stabilisce due ulteriori passaggi "amministrativi".

# Equo compenso

In primo luogo, dovrà essere stabilito l'ammontare dell'"equo compenso" che i titolari dei diritti avranno diritto a reclamare nel momento in cui ricomparissero: uno degli aspetti piú controversi della direttiva. Non è stata seguita la raccomandazione AIB riportata da Rosa Maiello (*op. cit.*), che il compenso fosse dovuto «solo in caso di di dimostrabile e significativo pregiudizio subíto», ma è ancora possibile stabilire un metodo di calcolo conforme agli obiettivi della legge. A nostro avviso, il compenso dovrà soddisfare alle seguenti condizioni:

- predeterminabile dagli utenti al momento dell'uso dell'opera, in quanto rischi legali futuri imprevedibili sono precisamente ciò che impedisce l'uso delle opere orfane;
- non proporzionale alla distribuzione che l'opera "liberata" riceve, altrimenti si penalizzerebbe chi meglio assolve al proprio dovere di massimizzare l'utilità pubblica dell'eccezione;
- e ovviamente contenuto, per non essere insostenibile.

Come punto di riferimento, il compenso potrebbe essere comparabile ai (già notevoli) costi sostenuti dall'ente per la ricerca diligente e la digitalizzazione, cosicché il costo di un programma di digitalizzazione non salga di un ordine di grandezza. Il titolare dei diritti dovrebbe inoltre "risarcire" tali costi qualora, nel porre fine allo stato di opera orfana, non consentisse all'ente in questione di continuare a far uso della scansione, cioè qualora operasse di fatto un "sequestro" dell'opera e scansione relativa. Il debito e il credito dell'ente potrebbero cosí, piú o meno, compensarsi. In caso contrario, si produrrebbe un perverso incentivo a che i titolari dei diritti si rendano irreperibili allo scopo di scaricare sulla collettività i costi di digitalizzazione per poi goderne tutti i benefici.

---

20  Laura Montanari. *Finiti i soldi per Google Books: la biblioteca digitale ferma a metà dell'opera*. «la Repubblica», 2014-07-11. <http://ricerca.repubblica.it/repubblica/archivio/repubblica/2014/07/11/finiti-i-soldi-per-google-books-la-biblioteca-digitale-ferma-a-meta-delloperaFirenze06.html>.

In effetti, proviamo ad adottare una similitudine che ci viene dagli USA: se le opere orfane sono piú correttamente denominate opere ostaggio, come indicato da Lydia Loren,[21] allora l'eccezione prevista dalla norma è un'ora d'aria a caro prezzo, e l'eventuale "equo compenso" un riscatto moralmente discutibile quanto quelli che finanziano iniziative terroristiche.[22]

Il cosiddetto "contributo per la copia privata", soprannominato "iniquo compenso"[23] da associazioni di consumatori e utenti, offre inoltre l'esempio di due criteri da *non* imitare (a meno di delegittimare ulteriormente, presso il pubblico, la legge sul diritto d'autore e lo stesso termine "equo compenso"):

- ha un carattere di imposta indiscriminata, indipendente dall'effettivo uso (della possibilità di copia);[24]
- le somme sono incamerate da un ente intermediario e prevalentemente consumate in costi o non distribuite.[25]

Ci auguriamo che il MIBACT impari dagli errori del passato.

## Registro

Sono stati stanziati 20 000 € per la costituzione del registro delle opere orfane, che essenzialmente dovrà quindi essere a costo zero. Auspichiamo che questa previsione sia compatibile con un ampio utilizzo dell'eccezione e che non si creino strutture ridondanti o processi burocratici complessi che renderebbero ancora piú costoso usufruirne.

In particolare, poiché i beneficiari diretti sono prevalentemente biblioteche, sembrerebbe ovvio integrare lo stato di opera orfana fra i metadati dell'OPAC SBN e schede di autorità connesse, che potrebbero quindi fungere da registro nazionale e automaticamente "inoltrare" l'informazione al registro europeo. Il dato dovrebbe ovviamente essere pubblicamente visibile e cercabile, nonché disponibile per uno scarico interoperabile.

In un passo successivo, l'informazione sullo stato di opera orfana potrebbe essere integrata con VIAF (e a cascata ARROW) e altri, riducendo i costi della ricerca diligente: un solo strumento assolverebbe a piú funzioni previste dalla norma. L'importanza di uno strumento unificato e di facile uso è dimostrata dall'esempio del Public Domain Calculator di Europeana.[26]

È necessario essere in grado di diffondere efficacemente e prontamente lo stato di opera orfana, determinato da istituzioni italiane per opere pubblicate in Italia. È un dovere di reciprocità nei confronti delle istituzioni (e utenti) degli altri paesi europei, i quali porteranno per lo piú il peso delle ricerche diligenti per le rispettive opere; ma è anche essenziale per assicurare la massima diffusione possibile della cultura italiana cosí rimessa in circolo. È, cioè, un forte interesse sia del Paese sia degli enti in questione, per la promozione delle rispettive collezioni.

---

21  Lydia Pallas Loren. *Abandoning the Orphans: An Open Access Approach to Hostage Works*. «Berkeley Technology Law Journal», Vol. 27: 1431, 2012. <http://ssrn.com/abstract=2049685>.
22  Si consenta questa ardita similitudine per evidenziare un'altra singolare distanza tra i contesti statunitense ed europeo. A questo proposito: *Remarks of Under Secretary David Cohen at Chatham House on "Kidnapping for Ransom: The Growing Terrorist Financing Challenge"*. 2012-05-10. http://www.treasury.gov/press-center/press-releases/Pages/tg1726.aspx
23  Associazione Wikimedia Italia. *Comunicato di risposta alle 10 domande SIAE*. 2011. <http://wiki.wikimedia.it/w/index.php?title=Comunicato_di_risposta_alle_10_domande_SIAE&oldid=74412>.
24  Come dimostrato dallo studio MIBACT nel 2014. http://www.altroconsumo.it/organizzazione/in-azione/azioni-in-corso/equo%20compenso%20-%20accesso%20agli%20atti
25  SIAE. *Rendiconto di gestione 2013*. <http://www.siae.it/documents/Siae_Documentazione_BILANCIOSIAE2013.pdf> (I debiti per mancata ridistribuzione del compenso per copia privata sono in crescita.)
26  http://www.outofcopyright.eu/

# Nuove possibilità

Alla luce di quanto sopra, che cosa possono e devono fare le biblioteche e altri enti beneficiari?

# Potenziare i servizi al pubblico

A tutt'oggi, sono pochissime le biblioteche che offrono servizi di riproduzione delle opere nelle proprie collezioni. In altri paesi europei, ogni biblioteca di una certa dimensione offre – per fare un esempio banale – fotocopiatrici ove l'utente può produrre un scansione e scaricarsene una copia. In Italia, invece, le fotocopiatrici sono inesistenti o dalle funzioni limitatissime e la riproduzione è vista come una minaccia da combattere. È necessario invertire questo approccio, dal momento che per la stragrande maggioranza delle collezioni (milioni di titoli) l'urgenza nazionale non è certo l'eccessiva riproduzione ma bensí la completa assenza di copie digitali.

Un esempio significativo è il servizio di riproduzione della Biblioteca Nazionale Centrale di Firenze.[27] Grazie alla collaborazione con una ditta specializzata, BNCF è in grado di offrire – a un costo molto competitivo – scansioni di qualità di una collezione ineguagliata in Italia: questo punto di forza va sviluppato, non mortificato. BNCF ha diritto a chiedere una tariffa maggiore agli editori, come già fa, anche per le opere orfane; anzi, non potrebbe rinunciare a tale diritto nemmeno se volesse (come invece ha fatto per le copie digitali di Google). Tuttavia, le due categorie "per motivi di studio" e "a scopo editoriale" usate per il servizio di BNCF sono superate; soprattutto, non è piú tollerabile l'applicazione indiscriminata del limite del 15 % per la riproduzione, né la minacciosa richiesta di un'autocertificazione per lo stato di "opera rara fuori dai cataloghi editoriali". Al contrario, dovrebbe essere data la massima visibilità al caso piú importante, quello delle opere nel pubblico dominio o non piú in commercio.[28]

Di piú: gli utenti dovrebbero essere sollecitati ad approfittare della possibilità di ottenere una copia digitale di alta qualità comodamente a casa propria e sul sito BNCF, nonché dell'onore di contribuire alla preservazione della cultura italiana. Le biblioteche del Paese, e in particolare le piú significative (quali le biblioteche nazionali, universitarie o centrali di comuni capoluogo), non hanno piú scusanti e dovrebbero tutte promuovere le nuove possibilità di digitalizzazione nei propri siti e OPAC. L'OPAC SBN dovrebbe mostrare la via, sollecitando gli utenti a effettuare ricerche diligenti per le opere di loro interesse e ad attivarsi per procacciare una digitalizzazione. Ciò potrebbe essere fatto in tutte le schede ove non sia certa l'impossibilità di procedere, ad esempio con un banale filtro "probabilistico" che escluda le opere pubblicate negli ultimi decenni o presenti in dozzine di biblioteche.

Non sono, queste, strade mai percorse prima: al contrario, ci sono illustri esempi internazionali. Il British Museum, ove non sia disponibile una riproduzione digitale di una sua opera, offre all'utente di farsi "sponsor of a reproduction" per il modico costo di circa 70 €.[29] 35 biblioteche di 12 paesi europei[30] partecipano al progetto "eBooks on demand" (EOD)[31] che dà la possibilità all'utente di richiedere una digitalizzazione e addirittura una ristampa di un'opera, a partire dal meta-OPAC (Ill. 4) o dall'OPAC delle singole istituzioni partecipanti. Per quanto ancora l'utente di lingua italiana potrà solo rivolgersi alle biblioteche svizzere per ottenere servizi moderni?

---

27  http://www.bncf.firenze.sbn.it/pagina.php?id=79&rigamenu=Riproduzioni
28  L'OPAC SBN riporta per la BNCF (nonostante l'incompletezza), oltre 560 000 schede per opere fino al 1923, che sono probabilmente nel pubblico dominio almeno in un paese al mondo; basta arrivare al 1975 per raggiungere circa 1,2 milioni di opere, ovvero la metà della consistenza del catalogo.

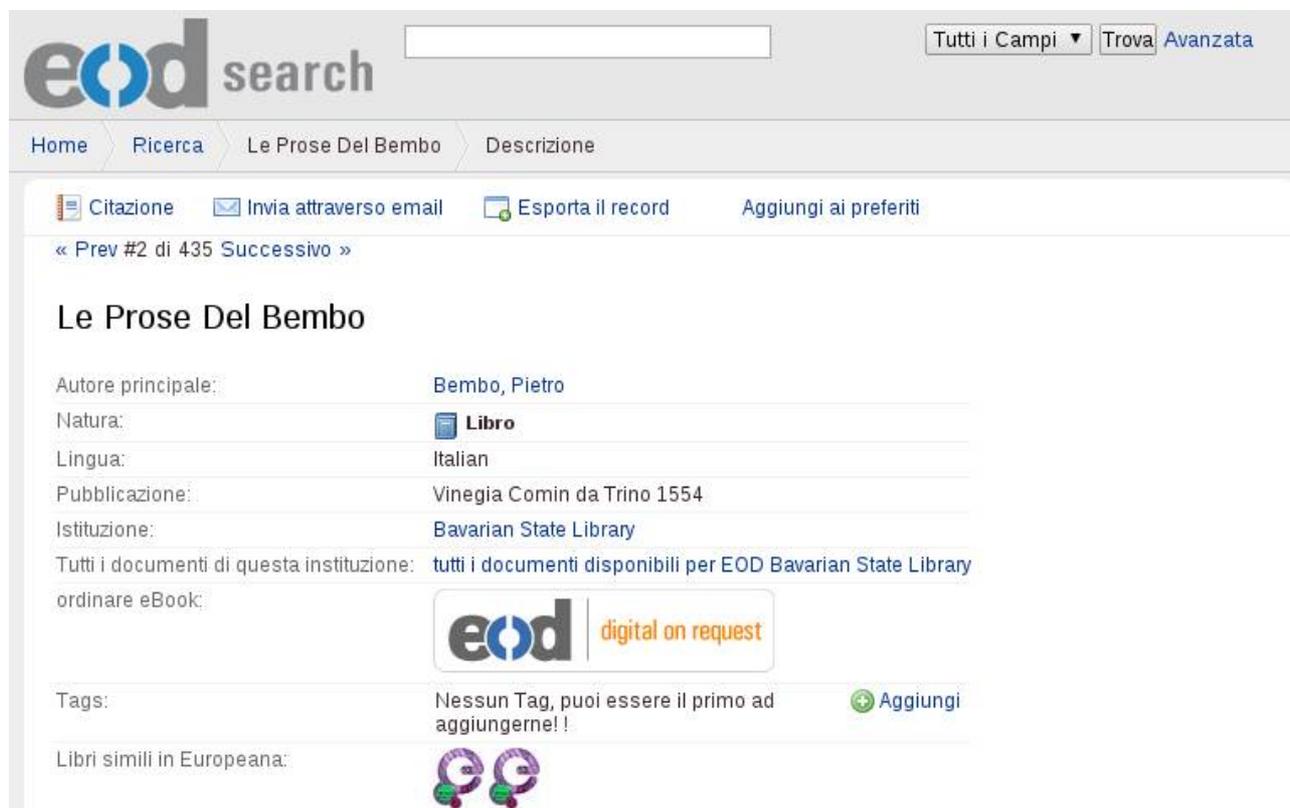

*Illustrazione 4: Un esempio di scheda del catalogo EOD*

## Strumenti disponibili: Internet Archive

Per ottenere risultati concreti nonostante i modestissimi investimenti, sia finanziari sia normativi, che lo stato italiano ha operato, è essenziale non solo coinvolgere la cittadinanza ma anche essere efficienti. Tranne un paio nel mondo, anche le biblioteche nazionali sono minuscole nel panorama delle biblioteche digitali globali, perciò l'unica via possibile è quella di partecipare a piattaforme condivise (e tendenzialmente a software libero). L'esempio più significativo è certamente l'Internet Archive.

Internet Archive (IA), fondazione non lucrativa statunitense nota principalmente per il suo servizio "Wayback machine" che archivia l'intero World wide web,[32] è in effetti una delle principali biblioteche digitali al mondo. Infatti ospita milioni di libri interamente accessibili in linea, di cui circa 2 milioni digitalizzati in proprio, con oltre 15 milioni di accessi al mese[33] e 300 000 utenti unici al giorno per il solo sito openlibrary.org.[34] Oltre 200 000 di questi libri provengono da biblioteche europee, numero che però impallidisce al confronto dei quasi 3 milioni da biblioteche statunitensi e canadesi.[35]

---

29  Petri Grischka. *The Public Domain vs. the Museum: The Limits of Copyright and Reproductions of Two-dimensional Works of Art*. «Journal of Conservation and Museum Studies», 12(1):8, 2014. <http://dx.doi.org/10.5334/jcms.1021217>.
30  http://books2ebooks.eu/it/content/la-rete-europea-delle-biblioteche
31  Günter Mühlberger – Silvia Gstrein. *EBooks on Demand (EOD): a European Digitization Service*. «IFLA Journal», 35. 1 (2009): 35-43. <http://archive.ifla.org/V/iflaj/IFLA-Journal-1-2009.pdf>.
32  http://web.archive.org/
33  http://www.the-digital-reader.com/2013/07/09/internet-archive-now-hosts-4-4-million-ebooks-sees-15-million-ebooks-downloaded-each-month/

IA è autofinanziato con donazioni (prevalentemente del suo fondatore e filantropo Brewster Kahle), quindi le biblioteche che ne richiedono i servizi di digitalizzazione devono trovare esse stesse i fondi necessari, che però sono contenuti grazie all'efficienza della struttura: indicativamente 10 centesimi di dollaro per pagina digitalizzata (contro 1 dollaro considerato standard). Peraltro IA ha fatto spesso ricorso a forza lavoro definibile di servizio civile o lavori socialmente utili, quindi un'istituzione pubblica può immaginare un vasto spettro di collaborazioni.

Nel caso piú semplice, però, non è necessario nulla di tutto ciò. IA è una biblioteca digitale partecipata: chiunque può caricare qualsiasi materiale, di qualsiasi dimensione, che desideri archiviare. Che siano 100 KB o 100 GB, un libro o centomila, chiunque al mondo può facilmente caricarli nel sito e vederli resi accessibili liberamente per sempre (salvo successive rimozioni per ragioni legali) e preservati a lungo termine. Non solo: l'infrastruttura procede automaticamente all'OCR[36] e alla conversione in una serie di formati aperti, fra cui il piú importante è il DjVu, che poi consente di sfogliare i libri con un software libero appositamente sviluppato da IA (Ill. 5).

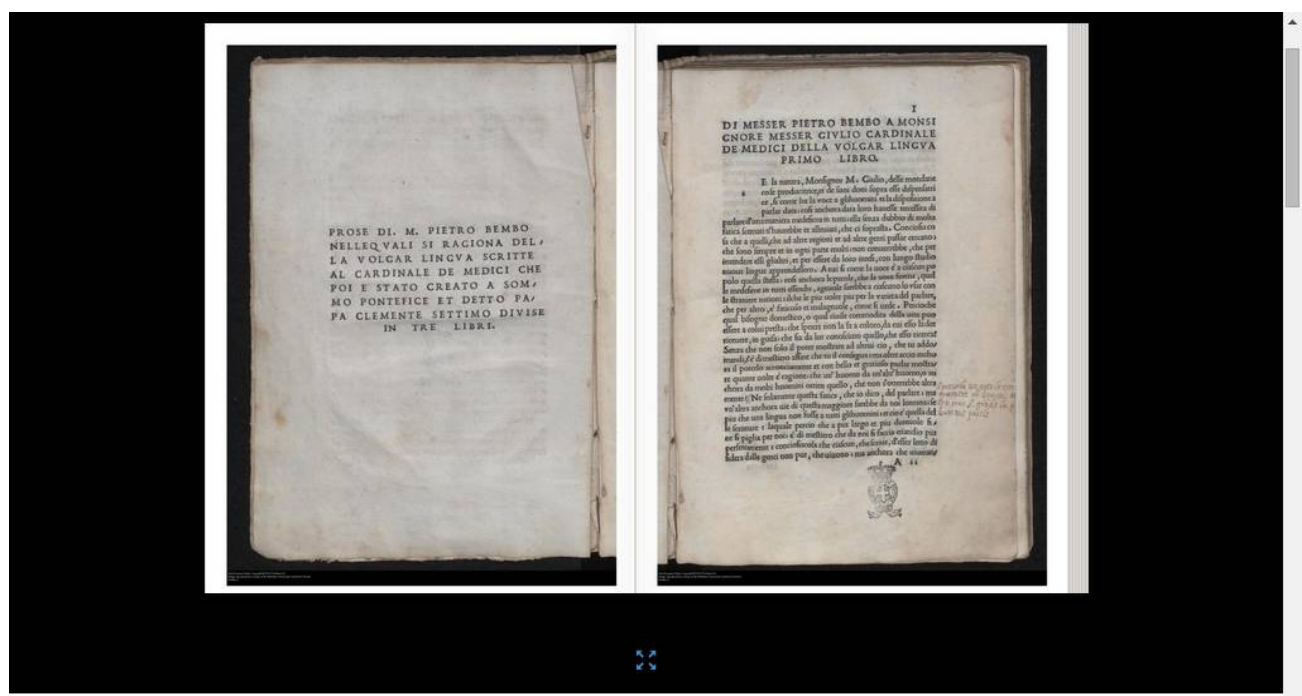

*Illustrazione 5: Pagina dei dettagli di un'opera in Internet Archive (v2), proveniente da BNCF*

---

34   https://archive.org/stats/
35   https://archive.org/details/texts
36   Anche se resta aperto il problema di come "allenare" l'OCR a specifici caratteri e alle centinaia di lingue non note agli attori commerciali. Le lingue di centinaia di milioni di persone al mondo non hanno OCR, e anche le altre hanno un tasso d'errore tuttora inaccettabile, che rende la trascrizione dei testi un compito immane tentato da pochissimi: PGDP, Wikisource, alcuni giornali.

IA ha già effettuato collaborazioni con biblioteche italiane: per esempio, BEIC (Biblioteca Europea di Informazione e Cultura) ha ricevuto da IA circa mille opere che desiderava includere nelle proprie collezioni, restituendo metadati di qualità che IA ha integrato nel proprio sistema.[37]

Può, dunque, essere la soluzione anche nel caso delle opere orfane? I dettagli legali sarebbero da verificare, ma la risposta si direbbe positiva, perché l'oggetto sociale di IA è precisamente l'interesse pubblico perseguito dall'eccezione per le opere orfane: a differenza di progetti piú complessi come il già visto Google Books, sembra facilmente adattabile ai requisiti di legge. Non solo beneficia del *fair use* garantito dalla sede statunitense, ma potenzialmente rientra fra gli enti beneficiari dell'eccezione per le opere orfane europee (purché le abbia nella propria collezione?[38]). In un'eventuale collaborazione, non impone alcuna restrizione alle biblioteche e non è necessario cedergli particolari diritti di utilizzo. Il suo uso potrebbe forse essere inquadrato in un mero contratto d'appalto, come a un agnostico fornitore di hosting, poi il visualizzatore (Ill. 6) potrebbe essere incluso direttamente nel sito della biblioteca (pratica che IA incoraggia). Usare il sito archive.org non richiede nessuna burocrazia, perciò anche la piú piccola iniziativa di digitalizzazione può trarne beneficio immediatamente, senza sprecare risorse in frizioni.

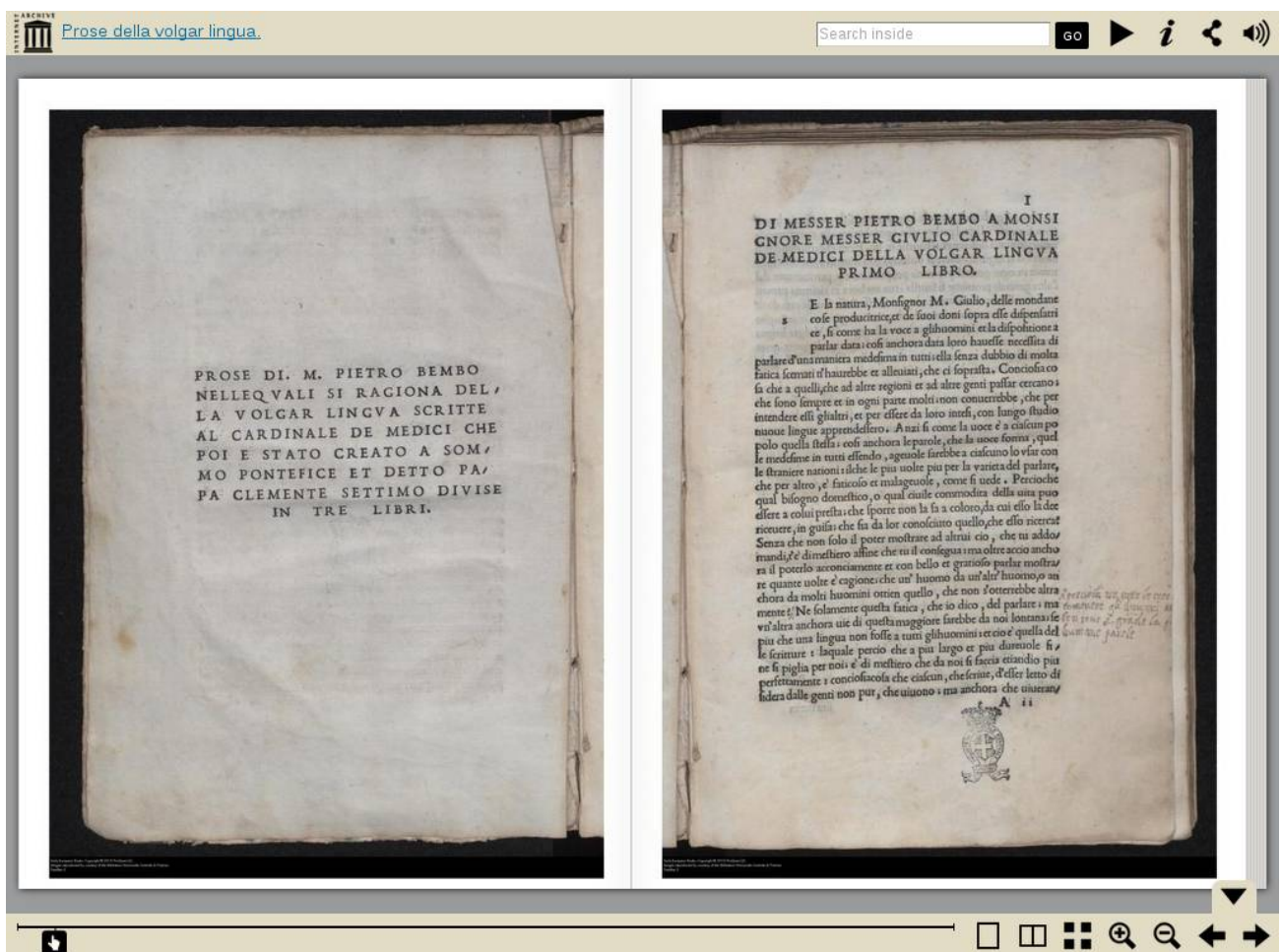

*Illustrazione 6: Il visualizzatore di libri di Internet Archive*

---

37  http://digitale.beic.it/primo_library/libweb/action/search.do?vl(freeText0)=internet+archive&fn=search&vid=beic
38  Del resto IA possiede anche un archivio fisico di centinaia di migliaia di opere altrimenti destinate al macero. http://blog.archive.org/2011/06/06/why-preserve-books-the-new-physical-archive-of-the-internet-archive/ In un panorama di finanziamenti pubblici decrescenti e biblioteche che chiudono, è prevedibile che in futuro possa acquisire fisicamente anche consistenti collezioni europee.

# Problemi ancora aperti

## Il pubblico dominio è molto di piú

È necessario ricordare che l'eccezione è solo un debole simulacro delle garanzie offerte dal pubblico dominio. In senso proprio, infatti, citando dal *Manifesto del pubblico dominio*,[39] il pubblico dominio è composto di «1. Opere d'autore per le quali sono scaduti i termini della tutela del copyright. [...] 2. Il bene comune ed essenziale dell'informazione che non è coperto da copyright».

La lunghezza dei termini, però, genera una contraddizione irrisolvibile. Il diritto d'autore è attribuito a caduchi esseri umani, ma ha una durata disumana: con 80 anni di vita dell'autore dopo la pubblicazione e 70 di durata dopo la morte si arriva già a 150 anni; quanti progetti individuali durano tanto? È ragionevole dunque che, molto prima di questo termine, qualcun altro debba subentrare a prendersi cura delle opere: candidate naturali sono le istituzioni che preservano e promuovono nei secoli la conoscenza. Tale conoscenza, infatti, fa parte del pubblico dominio in senso largo, inteso come «la materia grezza dalla quale viene ricavata la nuova conoscenza e si creano nuove opere culturali», per citare ancora il *Manifesto del pubblico dominio*. Ebbene, il primo principio affermato dal Manifesto è che «Il pubblico dominio è la regola, il copyright è l'eccezione»: per questo è legittimo che le opere orfane tornino anticipatamente un po' piú vicino al pubblico dominio, riducendo il divario fra pubblico dominio in senso largo e in senso stretto.

L'eccezione consente agli enti di perseguire i propri fini istituzionali in un regime di legalità, ma da sola non risolve i rischi legali connessi: si spera che il regime di "equo compenso" funzioni e il ricorso all'autorità giudiziaria sia residuale o inesistente, ma per quanto remota la possibilità porterà sempre gli enti a essere cauti. Inoltre, come ormai tutti danno per scontato, è molto improbabile che si assista a programmi centralizzati di digitalizzazione di massa delle opere orfane, fosse anche solo per i costi intrinseci della ricerca diligente.[40]

I rischi legali sono la faccia piú minacciosa, ma non la piú dannosa, del problema generale, cioè che le opere orfane non ritornano al pubblico dominio cui appartengono naturalmente: al contrario, restano in ostaggio ai loro inefficaci titolari, in un buio limbo solo parzialmente illuminato dalla nuova eccezione. Di conseguenza, le opere orfane non possono essere incluse in opere sotto licenza libera[41] né avere per esempio opere derivate quale una traduzione. Questa mancanza di libertà abbandona alla frustrazione non solo chi lavora negli spazi lasciati liberi dal copyright, come le biblioteche; ma anche chi coltiva il bene comune della cultura garantito dalle licenze libere o copyleft, come Wikipedia e i progetti Wikimedia.[42]

La frustrazione e l'inattività non portano però alcun bene: i beneficiari dell'eccezione hanno l'occasione di raccogliere le forze di tutti per perseguire l'obiettivo comune del pubblico dominio, che è l'unica garanzia per la promozione della conoscenza.

## Valorizzare il pubblico dominio

Non si può aspettare oltre un secolo prima di cominciare a coltivare il pubblico dominio di domani. Mentre noi aspettiamo, le opere di lingua inglese, o comunque selezionate da biblioteche statunitensi o di lingua inglese, dominano il panorama della cultura disponibile in Rete. Quelle, quindi, e non la cultura europea, forniscono piú largamente i mattoni per la costruzione della cultura contemporanea e futura, anche europea. Se le opere creative europee sono condannate al disuso oggi, domani saranno dimenticate; se per un'intera generazione, potrebbero essere marginalizzate per sempre (o almeno per secoli).

---

39  http://publicdomainmanifesto.org/italian
40  Basti pensare che persino nel progetto MIBAC con Google, riguardante opere nel pubblico dominio e quindi senza interferenza di terzi, i costi di identificazione e raccolta (catalogazione, selezione, collocazione e logistica) sono stati forse prevalenti: poiché erano in carico alla parte pubblica, i fondi sono presto finiti, mettendo a rischio la conclusione del progetto.
41  http://freedomdefined.org/Definition
42  https://www.wikimedia.org

Per cominciare, prima di poter rivendicare maggior libertà nell'uso delle opere culturali, è necessario rispettare quelle altrui. In particolare, come afferma lo *Statuto del pubblico dominio* di Europeana,[43] «Cio che è nel pubblico dominio deve rimanere nel pubblico dominio [...] L'utente legittimo di una copia digitale di un'opera nel pubblico dominio dovrebbe essere libero di (ri)utilizzare, copiare e modificare l'opera». Se fosse possibile per qualcuno appropriarsi le opere di altri, o già appartenenti a tutti, allora ogni sforzo sarebbe vano. Il *Manifesto del pubblico dominio* contiene ben cinque ulteriori raccomandazioni in questo senso (2, 4, 5, 7, 8), cui rimandiamo.

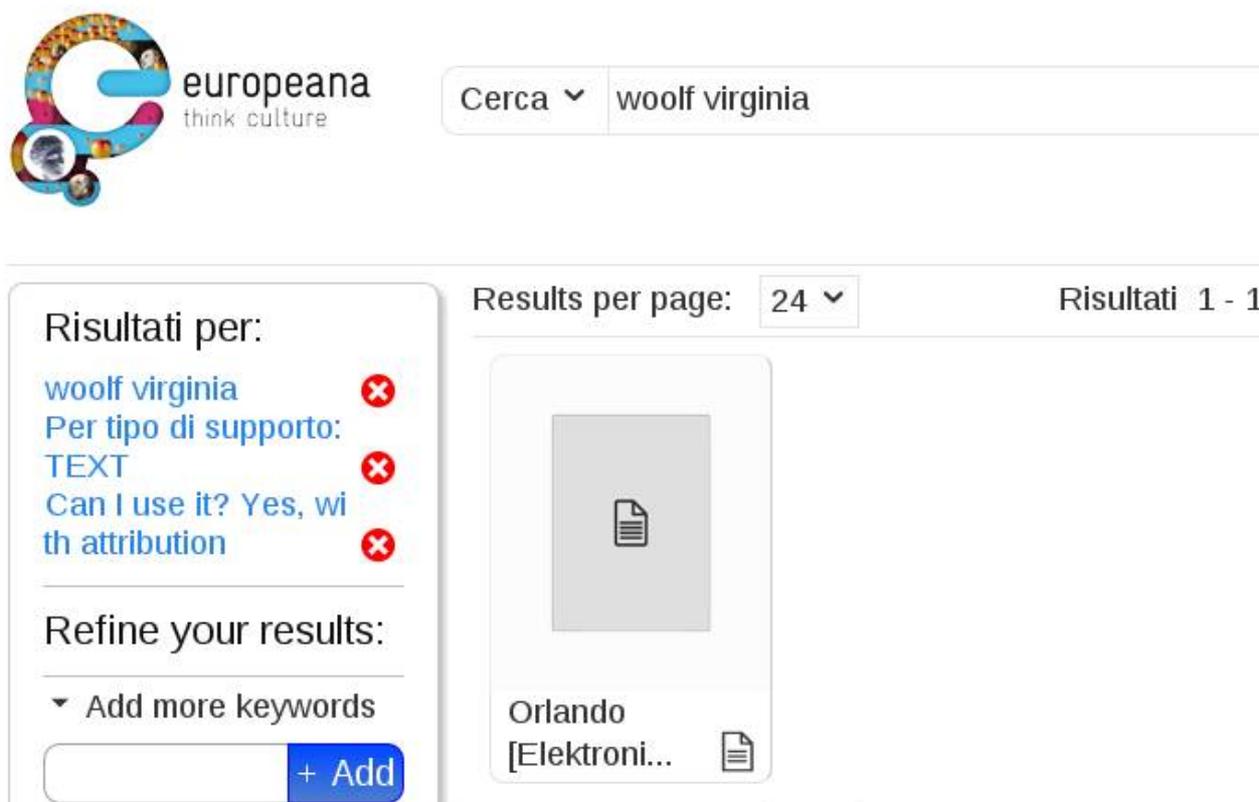

*Illustrazione 7: Le opere di Virginia Woolf sono in PD dal 2012, ma solo una lo è in Europeana*

Garantito ciò, la priorità è l'accorciamento dei termini del diritto d'autore (raccomandazione 1). La ricerca empirica dimostra che le opere musicali nel pubblico dominio raggiungono livelli superiori di digitalizzazione, e che negli USA i libri in pubblico dominio (ad esempio degli anni 1910) vendono di piú e soddisfano meglio la domanda che non quelli sottoposti alla restrizione monopolistica del diritto d'autore (come ad esempio quelli degli anni 1920 e 1930).[44] Ricordiamo che in origine la protezione era di 14+14 anni, nello Statute of Anne del 1710 poi ripreso negli USA: il prolungamento era condizionato a un'espressione di interesse da parte del titolare.

Senza entrare troppo nei dettagli, riteniamo che la direttiva sulle opere orfane tracci la via da seguire in futuro, perché il registro delle opere orfane afferma due strumenti che, se sviluppati, potrebbero avere notevoli risultati con piccolo sforzo.

---

43  http://pro.europeana.eu/c/document_library/get_file?uuid=d542819d-d169-4240-9247-f96749113eaa&groupId=10602
44  Paul J. Heald. *The Public Domain.* 2013-12-03. In: *The Law and Economics of Copyright*, a cura di Richard Watt. Routledge, 2014. <http://ssrn.com/abstract=2362983>.

- La necessità, per i titolari dei diritti, di dare un segno di vita e interesse per beneficiare della massima lunghezza del diritto d'autore.

- L'innalzamento dei diritti degli utenti su un'opera, almeno al livello concesso dal paese d'origine[45] (o, nel caso dell'UE, dallo stato membro piú permissivo?).

## Una seria riforma europea

Creatività, libertà di parola, democrazia, competizione e innovazione non possono esistere senza il pubblico dominio,[46] che è un «bene comune culturale, come l'aria, l'acqua, le foreste».[47] Persino WIPO ha indicato la necessità di rafforzare il pubblico dominio,[48] eppure sono molti gli ostacoli che si incontrano nell'accedervi;[49] per giunta il 47 % delle opere nel pubblico dominio non sono riconosciute come tali ove si usi il discrimine dei 140 anni dalla pubblicazione,[50] l'eliminazione retroattiva del pubblico dominio è una continua minaccia[51] e ulteriori norme fanno talvolta sparire interamente il pubblico dominio.[52]

Inoltre, nel 2014, la consultazione europea sul diritto d'autore ha dimostrato che la normativa è completamente sbilanciata a favore dei titolari dei diritti e a sfavore dei cittadini e che quindi una radicale riforma è necessaria.[53] La Commissione europea ha lavorato sulla base di false premesse per cui i titolari attuali dei diritti producono tutto il "valore" e gli altri sono parassiti;[54] sappiamo viceversa che la maggior parte della cultura risiede nel pubblico dominio e che gli "utenti" sono autori quanto chi pubblica in canali tradizionali.

Il commissario Oettinger sarà probabilmente al vertice del processo di riforma del diritto d'autore ed è in atto una ristrutturazione dei "ministeri" (direttorati) europei[55] potenzialmente positiva; ma non c'è ancora chiarezza sugli esatti confini delle competenze nella nuova Commissione, ancora meno sulle intenzioni concrete.

---

45  Questa è la raccomandazione 3 del Manifesto per il pubblico dominio ed è coerente con la Convenzione di Berna, che all'articolo 7.8 stabilisce la cosiddetta "regola dei termini piú corti" (*rule of the shorter term*).
<http://www.interlex.it/testi/convberna.htm#7>
46  Leonhard Dobusch. *The Digital Public Domain: Relevance and Regulation*. 2011-11-15. «HIIG Discussion Paper Series», No. 2012-02. <http://dx.doi.org/10.2139/ssrn.2011815>.
47  Melanie Dulong De Rosnay – Juan Carlos De Martin (a cura di). *The Digital Public Domain: Foundations for an Open Culture*. 2012. <http://porto.polito.it/2504638/>.
48  Cfr. Severine Dusollier., *op. cit.*
49  Randal C. Picker. *Access and the Public Domain.* 2013-02-08. «University of Chicago Institute for Law & Economics Olin Research Papers», No. 631. <http://dx.doi.org/10.2139/ssrn.2214176>.
50  Alex Clark – Brenda Chawner. *Enclosing the public domain: The restriction of public domain books in a digital environment*. «First Monday», [S.l.], may. 2014. ISSN 13960466. <http://journals.uic.edu/ojs/index.php/fm/article/view/4975>. doi:10.5210/fm.v19i6.4975.
51  Jennifer Jenkins. *In Ambiguous Battle: The Promise (And Pathos) Of Public Domain Day.* «Duke Law & Technology Review», 1-24 (2013). <http://scholarship.law.duke.edu/dltr/vol12/iss1/1>.
52  Federico Morando. *Diritti sui beni culturali e licenze libere (ovvero, di come un decreto ministeriale può far sparire il pubblico dominio in un paese)*. «Quaderni del Centro Studi Magna Grecia», Università degli Studi di Napoli, Federico II, 2011. <http://ssrn.com/abstract=2148343>.
53  Leonhard Dobusch. *Public hearing of the European Parliament's Committee on Legal Affairs and the Committee on Culture and Education on "The Future Development of Copyright in Europe"*. Brussels, 2014-11-11.
<http://governancexborders.com/2014/11/11/re-balancing-copyright-insights-from-the-eu-consultation/>
54  Paul Keller (2014). http://www.communia-association.org/2014/06/25/leaked-draft-of-commission-copyright-white-paper-based-on-flawed-assumptions/
55  LSE Media Policy Project. *The European Commission's new digital agenda duo*.
<http://blogs.lse.ac.uk/mediapolicyproject/2014/11/03/the-european-commissions-new-digital-agenda-duo/>.

# Conclusione

La direttiva sulle opere orfane è un fallimento largamente annunciato rispetto agli obiettivi dichiarati,[56] cosí come il "memorandum of understanding" fra titolari dei diritti e riutilizzatori, rimasto lettera morta.[57] A maggior ragione, è imperativo lavorare per restaurare il bene comune del pubblico dominio[58] ed è necessario che di questo compito si faccia carico chiunque abbia uno spazio di manovra o delle responsabilità nell'attuale quadro normativo. Biblioteche, musei e archivi sono messi alla prova dalla legge: devono usare tutto il proprio impegno e coraggio, ma possono ora dimostrare le opportunità insite nel rafforzamento del pubblico dominio, che è alle fondamenta del loro lavoro e della cultura di tutti, mediante la propria capacità di servire fattivamente il bene pubblico con questo nuovo strumento.

L'alternativa all'azione è la sparizione della cultura europea.

---

56  Paul Keller (2012). http://www.communia-association.org/2012/06/25/orphan-works-compromise-fails-to-deliver/
57  Associazione Wikimedia Italia. *Consultazione europea sul diritto d'autore*. <http://wiki.wikimedia.it/wiki/Consultazione_europea_sul_diritto_d%27autore#4._Mass_digitisation>
58  Lawrence Lessig (2006). *Re-crafting a Public Domain*. «Yale Journal of Law & the Humanities», Vol. 18, Iss. 3, Article 4. <http://digitalcommons.law.yale.edu/yjlh/vol18/iss3/4>.